*Article*

# I Can't Go to Work Tomorrow! Work-Family Policies, Well-Being and Absenteeism

José Aurelio Medina-Garrido *[ID], José María Biedma-Ferrer[ID] and Jaime Sánchez-Ortiz[ID]

INDESS, Universidad de Cádiz, 11406 Jerez de la Frontera, Spain; josemaria.biedma@uca.es (J.M.B.-F.); jaime.sanchez@uca.es (J.S.-O.)
* Correspondence: joseaurelio.medina@uca.es

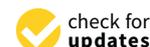

**Abstract:** Among the main causes of absenteeism are health problems, emotional problems, and inadequate work-family policies (WFP). This paper analyses the impact of the existence and accessibility of WFP on work absenteeism, by considering the mediating role of the well-being, which includes emotional as well as physical or health problems, that is generated by these policies. We differentiate between the existence of the WFP and its accessibility, as the mere existence of the WFP in an organisation is not enough. Additionally, workers must be able to access these policies easily and without retaliation of any kind. The model includes the hierarchy and the gender as moderating variables. To test the proposed hypotheses, a structural equation model based on the partial least squares structural equation modelling (PLS-SEM) approach is applied to a sample of employees in the service sector in Spain. On the one hand, the findings show that the existence of WFP has no direct effect on absenteeism; however, accessibility to these policies does have a direct effect on absenteeism. On the other hand, both the existence and accessibility of WFP have positive direct effects on emotional well-being. In addition, emotional well-being is positively related to physical well-being which, in turn, promotes a reduction in absenteeism. Finally, significant differences in the relationship between the existence of WFP and emotional well-being confirm the special difficulty of female managers in reconciling family life and work life.

## 1. Introduction

Absenteeism from work is a phenomenon that is currently of great interest to researchers and concern to managers [1]. According to the Addeco Group Institute Report [2], the absenteeism rate (understood as the ratio of hours not worked for occasional reasons to hours worked) reached a new high of 5.3% in Spain in 2018, the year with the highest rate of absenteeism in the period from 2000 to 2018. The analysis of the determining causes of such high levels of absenteeism is of particular interest to groups such as employers, researchers, and public administrations [3].

The increase in the rate of absenteeism has adverse socio-economic effects for the employee, the company, the regional economy, and the public administration. Previous studies state that continued absenteeism leads to the deterioration of skills that are required in the workplace and generates conflicts between colleagues [4]. Additionally, companies suffer reductions in productivity and competitiveness [5]. Finally, public administration experiences an increase in expenditure due to the social benefits that workers can claim [6].

Work absenteeism is a complex phenomenon that is influenced by various interrelated factors [7], which managers are aware of regarding their negative impact on labour costs and productivity [8]. However, managers experience great difficulties when they try to reduce work absenteeism [9,10].

Hence, managers must know the nature of the causes that affect work absenteeism so that they can properly manage them. Among the main causes of work absenteeism are emotional problems (lack of emotional well-being), health problems (lack of physical well-being), or inadequate work-family policies (WFP) [11–14]. The latter, in turn, can influence the well-being of workers [15–18], hence, WFP would also indirectly affect absenteeism. Therefore, we must consider that the well-being generated by these WFP has a mediating role in the WFP-absenteeism relationship. For this reason, organisations must assume an organisational, social and economic commitment to WFP [14,19–21], given their impact on absenteeism. In addition, we must consider that the welfare generated by WFP has a mediating role in the relationship between WFP and absenteeism. But, the mere existence of WFP in the organisation is a necessary but insufficient condition for workers to enjoy these policies. Further, workers should be able to access these policies without retaliation or any other inconvenience [22,23]. Considering the previous arguments, the aim of this paper is to analyse the impact of the existence and accessibility of WFP on work absenteeism. In addition, this paper also discusses the role of workers' emotional and physical well-being in the relationship between WFP and absenteeism.

To achieve the proposed objective, we applied a structural equation model that is based on the partial least squares structural equation modelling (PLS-SEM) approach to a sample of employees in the Spanish service sector. We separately analysed the accessibility and existence of WFP, which is a different approach compared to the previous research in the literature. Our findings show that the existence of WFP has no direct effect on absenteeism but, on the other hand, the accessibility of WFP does have a direct effect on absenteeism. Furthermore, both the existence and accessibility of WFP have positive direct effects on emotional well-being. In turn, emotional well-being and physical well-being act as mediators between WFP and absenteeism. Notwithstanding, we confirm the special difficulty of female managers in reconciling family life and work life.

This paper is structured as follows. Section two develops the theoretical model and the hypotheses that are to be tested. Section three describes the methodology that is developed, and we define the variables and their measurements, and apply a model of structural equations to the data that were collected. In section four, the results show that all of the hypotheses of our model are consistent with the postulated sign and all of the hypotheses are supported, except for the relation between the existence of WFP and absenteeism. In addition, we verify that the mediating effects in the model of emotional well-being and physical well-being are significant. In the final sections, we discuss and conclude the results of the study.

## 2. Background and Development of Hypotheses

*2.1. WFP and Absenteeism*

According to Role Theory, job satisfaction decreases when there is a conflict between work and family [24] and the results of several studies confirm this negative relationship empirically [25–27]. More specifically, the interference of work with respect to the family (versus interference of the family with respect to work) generates a decrease in the level of satisfaction of individuals in the organisation [28,29]. This type of conflict encourages employee absenteeism and also resignation from their post within the company [26,27]. Absenteeism, or not attending work when expected, could be defined as time lost on days a person had been away from work [30,31] due to reasons such as illness, family responsibilities, personal issues, or other reasons. Such work absenteeism is even more pronounced for employees who are married and with children [32]. Bansari [33] argues that a fear of losing their employment and a lack of potential opportunities will force workers to continue in their current jobs, but places them in unsatisfactory conditions that result in an increase in absenteeism levels in the workplace. Therefore, resolving the conflict and increasing job satisfaction would contribute to reducing or avoiding the likelihood that employees will be absent or even resign from their jobs [34].

If an organisation has adequate WFP, and those policies are accessible to employees, they will help to reduce family stress levels by increasing worker satisfaction levels and reducing work absenteeism [35–37]. WFP can be classified into four groups [22] (see Table 1): (1) Flexibility in working time (flexi-time); (2) flexibility through long paid and unpaid leave; (3) flexibility in work location (flexi-place); and (4) employee and family support services.

**Table 1.** Work-family policy (WFP) resources by groups.

| Groups | WFP |
|---|---|
| 1. Flexibility in working time | • Adapting the duration and distribution of working hours: continuous working day, breaks, and working time flexibility.<br>• Reduction in work hours to care for children and family members (part-time work).<br>• Compressed workweek.<br>• Taking holidays out of the regular vacation period.<br>• Breastfeeding leave.<br>• Other working time flexibility arrangements. |
| 2. Long paid and unpaid leave | • Leave to care for a hospitalised family member.<br>• Leave to take a family member to a health centre to receive medical assistance.<br>• Paid leave for sickness of a family member.<br>• Compressed breastfeeding leave (in days).<br>• Leave for international adoption.<br>• Leave to undergo a treatment of assisted reproduction.<br>• Unpaid leave (to care for children and dependent relatives).<br>• Leave for personal reasons.<br>• Unpaid additional holidays.<br>• Other paid and unpaid leave. |
| 3. Flexibility in the location of work | • Teleworking.<br>• Videoconferencing (except for teleworking).<br>• Transfer to a location nearer the family home. |
| 4. Employee and family support services | • Workplace nurseries.<br>• Childcare allowances.<br>• Allowances for employees with child or elder care responsibilities.<br>• Counselling on childcare services, schools, nursing homes for elderly and disabled people, etc.<br>• Work and family support services for employees and their families: psychological, legal, financial support, etc.<br>• Training in time and stress management.<br>• Counselling services on the WFP available.<br>• Other employee and family support services. |

In the specific case of Spain, the legislation establishes the minimum WFP that could be improved in sectoral agreements and private contracts. Among the WFP that are legislated in Spain are paid holidays (with a minimum of 30 calendar days per year), paid parental leave, and unpaid leave. Paid parental leave consists of maternity leave, paternity leave, breastfeeding leave, and leave for hazardous pregnancy or breastfeeding. Unpaid leave includes leave of absence and reductions in the working day to care for children or other family members who require it. Another noteworthy inclusion in Spanish legislation regarding WFP is the adaptation of the working day through greater time flexibility, shift changes, or even teleworking.

Cohen & Golan [32] argue that, even if there are WFP, the organisation may experience high work absenteeism if these policies are not properly implemented. Further, if labour policies are also not properly implemented in the company, then the conflict between work and family can lead to numerous negative consequences in the workplace, that is, certain counterproductive work behaviours for the organisation and its members, as is the case regarding work absenteeism [38].

As the above arguments suggest, although the existence of WFP can help reduce absenteeism, they are not enough and also studies in the literature often forget that these WFP must also be accessible to the worker. For this reason, for WFP to generate the expected benefits, workers must know of their existence [39–41]. However, workers must also perceive that WFP are accessible policies without reprisals or negative consequences for their career [42,43], otherwise they may not use WFP for fear of losing promotional opportunities, lack of commitment to the organisation, or even losing their job [23].

Since the mere existence of WFP is not enough [30], they should be studied separately in order to know if workers are aware of their existence and whether they perceive that WFP are accessible in practice.

In light of the above reasoning, the following hypotheses are proposed:

**H1:** *The existence of WFP is negatively related to work absenteeism.*

**H2:** *The accessibility of WFP is negatively related to work absenteeism.*

*2.2. WFP and Emotional Well-Being*

Studies in the literature analyse different types of well-being [44–46], among which workplace well-being stands out [47]. The study of occupational well-being pays special attention to psychological [45] or emotional well-being [46], in which pleasant and positive emotions prevail over negative emotions [48]. Emotional well-being describes how and why individuals experience their lives in a positive way [49]. High levels of emotional well-being imply more positive and less negative feelings [46]. Happiness and life satisfaction are components of emotional well-being, along with highly positive and less negative affectivity and mood [49]. Operationalising emotional well-being could be useful in measuring its impact on predicting job performance and employee retention [46].

The fact that workers experience emotions at work suggests that their experiences with their organisation's WFP will influence their emotional well-being [16]. Therefore, wherever the work-family conflict could have a negative impact on emotional well-being [15], WFP could be positively related to this well-being [22,23,50,51]. On the one hand, the work-family conflict generates psychological tension, which in turn generates a decrease in the emotional well-being of the worker at his job [52]. On the other hand, WFP will improve emotional well-being, since eliminating the conflict between the family and the workplace would have positive effects on people's satisfaction and happiness [53] and would reduce negative emotions [48].

In light of the above justification, and considering the previously argued difference between the existence and accessibility of WFP, we propose two hypotheses for study:

**H3:** *The existence of WFP is positively related to emotional well-being.*

**H4:** *The accessibility of WFP is positively related to emotional well-being.*

*2.3. Impact of Emotional Well-Being on Physical Well-Being*

According to the European Agency for Safety and Health at Work [54], psychosocial risks and work stress significantly affect people's health. According to the data of this Agency, half of the European workers think that work stress is common in their workplaces and that it causes them both mental and physical illnesses. Further, Gong et al. [55] affirm that there are deficiencies in the management of workers that can lead to psychological problems such as job stress, exhaustion, and depression.

Due to the high emotional pressures that workers experience and the prolonged duration of their workday, many of them lack physical activity [56], thus generating adverse effects such as musculoskeletal disorders [57], obesity [58], and diabetes [59].

Villalobos [60] notes that some of the diseases which are suffered by workers are caused by an inadequate business culture that leads the worker to workplace and personal stress. Martínez [61] notes that stress which is produced in the work environment is one of the sources that generates psychosocial and physical risks, thus causing a deterioration in both worker health and the work environment. In this sense, Barlow [62] identifies various sources of workplace stress that expose people to daily demands that can affect their mental and physical health and well-being.

From the above discussion, it can be considered that if there are no WFP, the existence of emotional well-being is hindered, which leads to increased stress and a reduction in the physical well-being of the worker. Therefore, we propose the following hypothesis:

**H5:** *Emotional well-being is positively related to physical well-being.*

*2.4. Physical Well-Being and Absenteeism*

It is indicated in the literature that when workers are not motivated, they increase the chances that they will suffer a physical deterioration in their health and that absenteeism will increase [63]. Proper WFP can increase motivation, increase the physical well-being of the employee, and ultimately reduce absenteeism.

Martínez-López & Saldarriaga-Franco [64] undertook a study to determine which were the most common illnesses that caused absenteeism, whereby they considered three types of individuals: physically active, sedentary, and mixed. According to that study, the main illnesses that caused absenteeism were respiratory, musculoskeletal, and trauma illnesses. Those individuals who practiced sports were less likely to suffer from the aforementioned diseases. However, sedentary individuals were more likely to suffer from diseases that caused them disability and subsequently generated absenteeism.

Viana et al. [65] state that there is an association between physical activity and worker performance. The physical well-being of the worker is one of the factors that contributes to the maintenance of good health in the workplace and, therefore, reduces the level of absenteeism considerably [66,67]. Furthermore, to reduce the level of absenteeism in organisations as a consequence of physical illnesses, it is necessary to develop certain awareness programmes that can generate incentives for workers to take up physical exercise programmes [68].

Therefore, we propose the following hypothesis:

**H6:** *Physical well-being is negatively related to absenteeism in the workplace.*

*2.5. Moderating Effect of Gender and Hierarchy*

Empirical evidence on the work-family interaction shows that men and women do not manage this duality of roles in the same way, and that women report more difficulties in making their work and family roles compatible [69,70]. In this sense, it is noted in the literature that the work-family conflict is greater in women than in men, due to the overload in caring for the family, specifically children and the elderly [71–73]. For this reason, one of the most current policies and measures in the labour market is aimed at reducing this phenomenon of labour turnover in women [74,75]. Boeckmann, Misra, & Budig [76] affirm that the balance between family life and work life is more complex in the case of women than of men, since women find it more difficult to disinhibit themselves from family problems and, more specifically, from children's problems. Various studies [77,78] affirm that the working woman carries a double burden as an employee, a housewife, and the main reference for family care.

These studies allow us to predict that the existence and accessibility of WFP will be more relevant for women and that, therefore, their impact on well-being and absenteeism may differ from that of men. In addition, the impact of WFPs may also differ depending on whether or not an individual is a manager in the organisation. In the case of female managers, being a woman and a manager is a double handicap, which implies a special difficulty in reconciling family life and work life [79–81]. However, there is also evidence that the differences between men and women may not be significant in all cases, due to the influence on this phenomenon of cultural and sector conditioning factors [22,23]. Given the relevance of this research question, the present paper analyses the moderating role that gender and the position occupied in the hierarchy (manager or employee) could have in the previously established hypotheses. Therefore, we propose the following hypothesis:

**H7:** *Managerial status moderates model relationships.*

**H8:** *Gender moderates model relationships.*

Considering the relationships outlined above in the hypotheses and the moderating effect discussed above, Figure 1 shows the theoretical model that is proposed for testing.

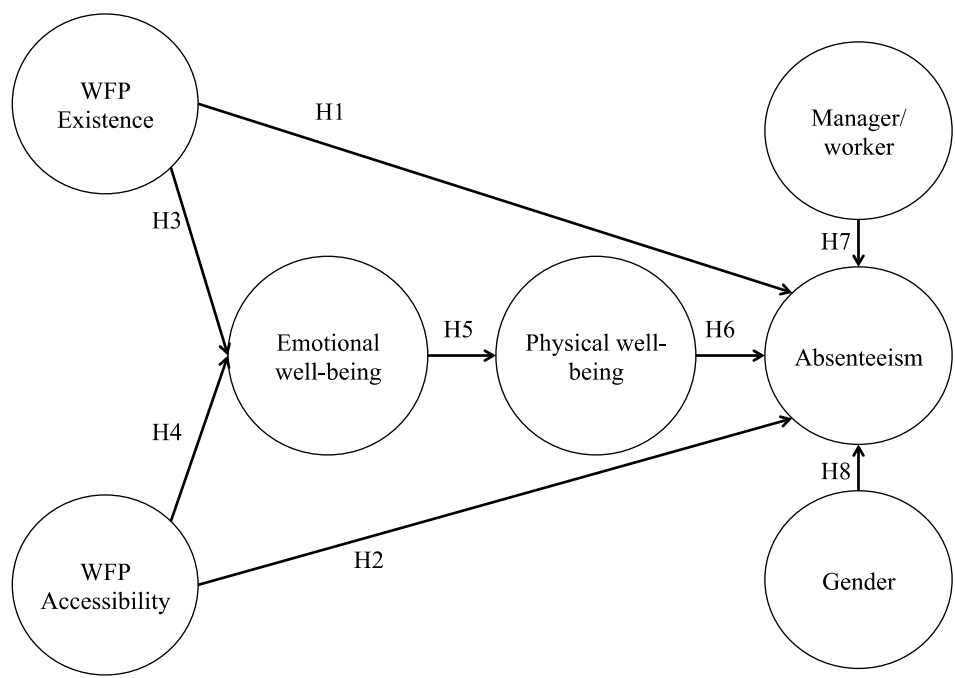

**Figure 1.** Theoretical model for WFP-well-being-absenteeism.

## 3. Methodology

*3.1. Sample*

The obtained sample consisted of 584 workers from the service sector in Spain, surveyed between January and November 2016. We have selected the service sector because it is a crucial sector in the dynamics of the Spanish economy and because it is the engine of employment in Spain [82]. This sector converges in the European countries of reference, with a higher weight of services and manufacturing, and an agricultural sector with reduced importance [83].

Applying the pre-evaluation of the data according to the guidelines proposed by Hair et al. [84], we eliminated 49 observations because they contained missing values for all of the indicators of a certain variable, and 11 observations because they contained a high percentage of missing data (more than 15%). The proportion of cases that were withdrawn from the original sample was 10.3%. Therefore, we carried out the study with 524 valid questionnaires. We analysed the results of a T-test of related samples to assess the possible differences that may occur in each of the variables. When applying the analysis, we observe that there has only been a significant change in the indicator "Because the usual babysitter was not available" of the variable "Absenteeism." Regarding the respondents, 45.2% were women, 16.6% were managers, more than half of the respondents had undertaken university studies, and the mean age was 44.7 years. Table 2 shows these and other characteristics of the sample.

In order to analyse the interaction effect of the gender and hierarchy variables, the sample was divided into four subsamples: (1) female managers, (2) female employees, (3) male managers, and (4) male employees, made up of 34, 203, 53, and 234 individuals, respectively.

Considering a statistical power of 0.8 and a default alpha level of 0.05, the sample (n = 524) permitted the detection of very small effect sizes that, according to Cohen [85], are more difficult to identify.

Table 2. Characteristics of the sample.

|  | % | Mean |
|---|---|---|
| Women | 45.2% |  |
| Managers | 16.6% |  |
| Age |  | 44.7 years old |
| Without a partner | 9.6% |  |
| With a partner and who also works | 71.0% |  |
| With children under 18 years old | 61.8% |  |
| With dependent relatives in the ascending line who need help | 23.5% |  |
| With disabled people in their care who need help | 5.2% |  |
| With university studies, master's or doctorate | 67.9% |  |
| With a university education | 56.3% |  |
| Years working in their organisation |  | 18.2 years |
| Hours of work per week |  | 39.4 h per week |

*3.2. Measurements*

We modelled the variables that are included in this study as compound events (CEs), since they are design variables or artefacts in which the indicators (linear combination) make up the variable. More specifically, we modelled all of the variables as Mode A compound events (i.e., using correlation weights) since the indicators were correlated [86].

We used multiple indicators that assessed the respondents' degree of agreement with various statements using a 7-point Likert scale. To measure the existence of WFP, we adapted the Families and Work Institute (FWI) scale [87,88] which consists of five items. The WFP accessibility variable was a combination of contributions in accordance with Anderson, Coffey, and Byerly [89] and the Families and Work Institute [87,88], consisting of fifteen items. It is essential to point out that both the existence and accessibility of WFP are perceptual variables, that is, their measurement depends on the subjective perception of workers. Both may differ from the actual WFP that exist in the organisation and their real accessibility. This perceptual character does not imply any bias in our model since it is the perceptions and not the reality that influence the well-being of the workers. We measured the emotional well-being variable by combining the contributions of the scales of Warr [90] and Kossek, Colquitt and Noe [91] on a reflective scale of fifteen items. To measure physical well-being, we used a nine-item scale from the work of Kossek, Colquitt and Noe [91]. We performed the measurement of absenteeism with a numerical scale that is composed of four items from the work of Anderson et al. [89], taking into account that the values which were established as "ten days or more" were assigned the value 10. Table 3 shows the items ordered by construct. To analyse the moderating effect, we formed four differentiated groups, which are: (1) female managers; (2) female employees; (3) male managers; and (4) male employees, while considering hierarchy and gender.

Table 3. Constructs and indicators.

| WFP existence (5 items) [87,88] |
|---|
| 1. Your organisation offers time off for family reasons.<br>2. Your organisation reports on time off for family reasons.<br>3. You know what time off for family reasons consist of.<br>4. You have used time off for family reasons.<br>5. You know employees who have used time off for family reasons. |

**Table 3.** *Cont*.

| |
|---|
| WFP accessibility (15 items) [87–89] |
| 1. It is hard for you to take time off during your workday to take care of personal or family matters. (Reverse-coded) * <br> 2. There is an unwritten rule at my place of employment that you cannot take care of family needs on company time. (Reverse-coded) <br> 3. At my place of employment, employees who put their family or personal needs ahead of their jobs are not looked upon favourably. (Reverse-coded) <br> 4. If you have a problem managing your work and family responsibilities, the attitude at my place of employment is: "You made your bed, now lie in it!" (Reverse-coded) <br> 5. If you are asked to work extra or overtime hours, you can refuse to work these extra hours without negative consequences at work. * <br> 6. At my place of employment, employees have to choose between advancing in their jobs or devoting attention to their family or personal lives. (Reverse-coded) <br> 7. Employees who ask for time off for family reasons or try to arrange different schedules or hours to meet their personal or family needs are less likely to get ahead in their jobs or careers. (Reverse-coded) <br> 8. I have the schedule flexibility I need at work to manage my personal and family responsibilities. * <br> Why did you choose not to request a flexible work arrangement? <br> 9. My job responsibilities do not allow it. (Reverse-coded) * <br> 10. There would be negative consequences for my job advancement. (Reverse-coded) <br> 11. There would be negative consequences for my current or future earnings. (Reverse-coded) <br> 12. My manager is not (would not be) supportive. (Reverse-coded) <br> 13. My co-workers are not (would not be) supportive. (Reverse-coded) <br> 14. It might mean that others at work would have more to do. (Reverse-coded) * <br> 15. It might make me look less committed to my job or career. (Reverse-coded) |
| Emotional well-being (15 items) [90,91] |
| Thinking of the past few weeks, how much of the time has your job made you feel each of the following? (Scale: never, very rarely, occasionally, some of the time, much of the time, most of the time, all of the time): <br> 1. Tense. (Reverse-coded) <br> 2. Uneasy. (Reverse-coded) <br> 3. Worried. (Reverse-coded) <br> 4. Calm <br> 5. Contented <br> 6. Relaxed <br> 7. Depressed. (Reverse-coded) <br> 8. Gloomy (Reverse-coded) <br> 9. Miserable. (Reverse-coded) <br> 10. Cheerful <br> 11. Enthusiastic <br> 12. Optimistic <br> 13. Angry. (Reverse-coded) <br> 14. Annoyed. (Reverse-coded) <br> 15. Irritated. (Reverse-coded) |
| Physical well-being (9 items) [91] |
| Thinking of the past few weeks, how much of the time has your job made you feel each of the following? (Scale: never, very rarely, occasionally, some of the time, much of the time, most of the time, all of the time): <br> 1. Your hands trembled enough to bother you. (Reverse-coded) <br> 2. You were bothered by shortness of breath when you were not working hard or exercising. (Reverse-coded) <br> 3. You were bothered by your heart beating hard. (Reverse-coded) <br> 4. You were bothered by your heart beating faster than usual. (Reverse-coded) <br> 5. Your hands sweated so much that you felt damp and clammy. (Reverse-coded) <br> 6. You had spells of dizziness. (Reverse-coded) <br> 7. You were bothered by having an upset stomach or stomachache. (Reverse-coded) <br> 8. You had a loss of appetite. (Reverse-coded) <br> 9. You had trouble sleeping at night. (Reverse-coded) |
| Absenteeism (4 items) [89] |
| How many days did you miss work in the past 3 months . . . <br> 1. to care for a sick child? <br> 2. because their usual child care was not available? <br> 3. for other family reasons. <br> 4. to carry out personal business. ** |

\* This item was removed in the assessment of the measurement model. ** We added the absenteeism item number four.

*3.3. Methodology*

To test the proposed hypotheses, we propose a structural equation model based on the PLS-SEM approach. We follow the methodological recommendations of Hair et al. [92] to obtain consistent estimates, since we have variables that are modelled as compound events (CEs). In the analysis, we used the path modelling software SmartPLS 3.2.7 [93] and we applied the criterion of substituting the mean for missing values in the data treatment.

The evaluation of the model was carried out according to the following stages: (1) assessment of the global model, (2) assessment of the measurement model, (3) assessment of the structural model, and finally (4) analysis of the moderating effect.

*3.4. Common Method Bias*

The common method bias (CMB) is a phenomenon that is caused by the measurement method which is used in an SEM study or by the way in which particular questions are answered in a survey [94]. To analyse this problem, we performed a full collinearity test based on variance inflation factors (VIFs), as proposed by Kock [94]. According to this test, a VIF value greater than 3.3 shows pathological collinearity, which would indicate that the model is contaminated by the bias of the common method. In our study, the maximum value is 2.084 (Table 4), hence, it can be considered that the model is free of bias.

**Table 4.** Full collinearity.

| Variables | Existence | Accessibility | Emotional Well-Being | Physical Well-Being | Absenteeism |
|---|---|---|---|---|---|
| Variance inflation factor (VIF) | 1.132 | 1.279 | 2.084 | 1.896 | 1.024 |

## 4. Results

*4.1. Assessment of the Global Model*

According to Henseler, Hubona, and Ray [95], the goodness of fit (GOF) of the global statistical model should be the starting point for the valuation of that model, and if it does not fit the data, then the estimates that are obtained may be meaningless and, consequently, the conclusions may be questionable.

In our study, we carried out two evaluations of the global model: (1) the first evaluation took into account all of the indicators of the model before evaluating the measurement model, and (2) the second evaluation was subsequent to the study of the measurement model and the elimination of indicators that do not meet the necessary requirements.

For the study of the global model, we used a measure of approximate model fit in accordance with [95]. More specifically, we analysed the standardised root mean square residual (SRMR), whose threshold is 0.8 [96]. The results gave a value of 0.066 before eliminating the indicators and a value of 0.051 after eliminating the indicators, hence, the model is better at eliminating indicators that present reliability and validity problems, and consequently, we can affirm that we have an approximately true model.

*4.2. Assessment of the Measurement Model*

Since the model is made up of Mode A estimated compound events, we analyse: (1) reliability, to verify that the measurement is made in a stable and consistent way, and (2) validity, to verify that the indicators accurately measure what we wish to measure

Firstly, we analyse the individual reliability of the item, examining the loads of the indicators with their respective constructs (or composites). We adopt the threshold proposed by Carmines and Zeller [97], in which items with loads greater than 0.707 are accepted. However, some researchers argue that items with loads in the range of 0.4–0.707 should not be eliminated if they do not pose problems for the rest of the stages of the measurement model. In our study, we initially eliminated three items which correspond to the variable WFP accessibility with very low loads (below 0.4). Subsequently, we eliminated two more items in this variable in order to meet the requirement for convergent validity, hence, we were able to increase the value of the average variance extracted (AVE).

Secondly, we analysed the reliability of the construct to determine whether the items, that are measures of their constructs (or composites), have similar scores. For this, we took into account the measures which correspond to the composite (or construct) reliability [98]. Nunnally and Bernstein [99] suggest values which are higher than 0.8 for advanced stages of research. As can be seen in Table 5, all constructs are above the threshold for both reliability measures.

Subsequently, we studied the convergent validity (within the larger construct validity) in order to verify that the indicators represent a single underlying construct. As a measure, we use the average variance extracted (AVE), and as can be observed in Table 5, the AVE values show that there is convergent validity when exceeding the recommended minimum value of 0.5 [100].

Finally, we checked the discriminant validity, that is, the degree to which a given construct is unrelated with other constructs, and Table 6 shows this verification using the Heterotrait-Monotrait (HTMT) ratio that was developed by Henseler et al. [101]. The discriminant validity can also be verified with the Fornell and Larcker [100] criterion by using the correlation matrix between variables.

Once the previously described items had been eliminated, the model met the necessary reliability and validity requirements (refer to the results in Tables 5 and 6).

**Table 5.** Results of the measurement model.

| Construct/Indicators | Weight | Composite Reliability | ρA | AVE |
|---|---|---|---|---|
| WFP existence (Composite, Mode A) | | 0.861 | 0.830 | 0.556 |
| 1. Your organisation offers them | 0.804 | | | |
| 2. Your organisation reports on them | 0.786 | | | |
| 3. You know what they consist of | 0.747 | | | |
| 4. You have ever used them | 0.597 | | | |
| 5. You know employees who have used them | 0.775 | | | |
| WFP accessibility (Composite, Mode A) | | 0.913 | 0.900 | 0.517 |
| 2. There is an unwritten rule | 0.611 | | | |
| 3. Not looked on favourably | 0.743 | | | |
| 4. "You made your bed, now lie in it!" | 0.762 | | | |
| 6. Employees have to choose | 0.741 | | | |
| 7. Less likely to get ahead in their jobs | 0.776 | | | |
| 10. Negative consequences for my job | 0.830 | | | |
| 11. Negative consequences for my earnings | 0.715 | | | |
| 12. My superior would not support it | 0.679 | | | |
| 13. My co-workers would not support it | 0.486 | | | |
| 15. Appear to be less committed | 0.783 | | | |
| Emotional well-being (Composite, Mode A) | | 0.957 | 0.955 | 0.596 |
| 1. Tense | 0.778 | | | |
| 2. Uneasy | 0.792 | | | |
| 3. Worried | 0.764 | | | |
| 4. Calm | 0.732 | | | |
| 5. Contented | 0.734 | | | |
| 6. Relaxed | 0.710 | | | |
| 7. Depressed | 0.817 | | | |
| 8. Gloomy | 0.836 | | | |
| 9. Miserable | 0.796 | | | |
| 10. Cheerful | 0.775 | | | |
| 11. Enthusiastic | 0.673 | | | |
| 12. Optimistic | 0.739 | | | |
| 13. Angry | 0.783 | | | |
| 14. Annoyed | 0.807 | | | |
| 15. Irritated | 0.830 | | | |

Table 5. *Cont.*

| | | | | |
|---|---|---|---|---|
| Physical well-being (Composite, Mode A) | | 0.950 | 0.942 | 0.677 |
| 1. Hands trembled | 0.770 | | | |
| 2. Shortness of breath | 0.869 | | | |
| 3. Heart beating hard | 0.899 | | | |
| 4. Heart beating faster | 0.891 | | | |
| 5. Hands sweated | 0.796 | | | |
| 6. Spells of dizziness | 0.812 | | | |
| 7. Stomachache | 0.826 | | | |
| 8. Loss of appetite | 0.786 | | | |
| 9. Trouble sleeping | 0.743 | | | |
| Absenteeism (Composite, Mode A) | | 0.887 | 0.902 | 0.664 |
| To care for a sick child | 0.828 | | | |
| Babysitter unavailable | 0.922 | | | |
| For other family reasons | 0.775 | | | |
| To carry out personal business | 0.721 | | | |

Note: ρA: Dijkstra-Henseler's rho, AVE: average variance extracted. Note: See Table 3 for completed items.

Table 6. Measurement model. Discriminant validity. HTMT Ratio (Heterotrait-Monotrait ratio).

| | Existence | Accessibility | Emotional Well-Being | Physical Well-Being | Absenteeism |
|---|---|---|---|---|---|
| Existence | | | | | |
| Accessibility | 0.199 | | | | |
| Emotional well-being | 0.311 | 0.406 | | | |
| Physical well-being | 0.297 | 0.284 | 0.708 | | |
| Absenteeism | 0.110 | 0.152 | 0.107 | 0.178 | |

*4.3. Structural Model*

For the evaluation of the structural model, we analysed the hypotheses raised in the model using the bootstrapping resampling technique with 5000 samples. In this way, we were able to assess the magnitude, sign, and significance of the relationships between the variables. Furthermore, we analysed the predictive power of the model using the coefficient of determination ($R^2$) of the endogenous variables and its subsequent decomposition of the explained variance. This allowed us to understand the importance of each of the antecedent variables in the dependent variable. Finally, we use the Cohen [85] rules to assess the size of the effects. These results are reflected in Table 7.

Table 7. Direct effects. Hypothesis.

| | Direct Effect | *p*-Value | *t*-Value | CI | Supported | Explained Variance | $f^2$ |
|---|---|---|---|---|---|---|---|
| Absenteeism ($R^2$ = 0.032) | | | | | | | |
| H1(−): Existence | −0.043 | 0.111 | 1.223 | (−0.091; 0.015) | No | 0.38% | 0.002 |
| H2(−): Accessibility | −0.061 | 0.040 | 1.749 | (−0.118; −0.009) | Yes | 0.63% | 0.004 |
| H6(−): Physical well-being | −0.134 | 0.027 | 1.928 | (−0.215; −0.006) | Yes | 2.16% | 0.016 |
| Emotional well-being ($R^2$ = 0.190) | | | | | | | |
| H3(+): Existence | 0.219 | 0.000 | 5.526 | (0.156; 0.286) | Yes | 6.09% | 0.057 |
| H4(+): Accessibility | 0.341 | 0.000 | 8.596 | (0.279; 0.410) | Yes | 12.92% | 0.139 |
| Physical well-being ($R^2$ = 0.473) | | | | | | | |
| H5(+): Emotional well-being | 0.688 | 0.000 | 32.547 | (0.654; 0.723) | | 47.33% | 0.898 |

The $R^2$ values that were obtained for this model show a low predictive power for the dependent variable emotional well-being, a moderate predictive power for the variable physical well-being and a very weak predictive power for the variable absenteeism.

All of the hypotheses are consistent with the postulated sign and all are supported except for H1, hence the existence of work-family reconciliation policies does not affect absenteeism in the workplace.

On the other hand, the rest of the hypotheses are fulfilled with a very small effect for H2 and H6, a small effect for H3 and H4, and a large effect for H5, according to the Cohen [85] tables.

Finally, we analysed the possible mediating effects of the model and checked that they are significant, so that we could deduce that the variables emotional well-being and *physical well-being* function as mediating variables within the proposed model (Table 8).

Table 8. Indirect effects. Mediation.

|  | Indirect Effect | *p*-Value | *t*-Value | CI | Supported |
|---|---|---|---|---|---|
| Existence→Emotional well-being→Physical well-being→Absenteeism | −0.020 | 0.040 | 1.754 | (−0.036; −0.001) | Yes |
| Accessibility→Emotional well-being→Physical well-being→Absenteeism | −0.031 | 0.034 | 1.826 | (−0.054; −0.001) | Yes |
| Existence→Emotional well-being→Physical well-being | 0.151 | 0.000 | 5.325 | (0.107; 0.200) | Yes |
| Accessibility→Emotional well-being→Physical well-being | 0.235 | 0.000 | 8.190 | (0.190; 0.284) | Yes |
| Emotional well-being→Physical well-being→Absenteeism | −0.092 | 0.028 | 1.917 | (−0.149; −0.004) | Yes |

*4.4. Moderating Effect*

Finally, we carried out a multi-group analysis (MGA) to test the interaction effect of the moderating (or dampening) variable that is associated with gender and hierarchy. To do this, we divided the sample into four groups: (1) female managers, (2) female employees, (3) male managers, and (4) male employees. First, we analysed the measurement invariance (MI) to ensure that the differences between groups were due to the path coefficients and not to the parameters of the measurement model. We verified the existence of partial invariance in one case, total invariance in another case, and in the remaining four cases the measurement invariance was not met (Table 9). Therefore, we could not apply the permutation-based analysis that was developed by Chin [102] to these latter cases, in order to assess whether there are significant differences between each pair of groups. We only performed the permutation-based analysis in the two cases where there was measurement invariance. Following the analysis, we found that there were significant differences only in the relationship between the existence of WFP and emotional well-being between the groups of "female managers" and "female employees" (Table 10). These findings only partially support H7 and H8.

Table 9. Results of the measurement invariance assessment (MICOM).

| | Step 1 | Step 2 | | | Step 3a | | | | Step 3b | | | | |
|---|---|---|---|---|---|---|---|---|---|---|---|---|---|
| | Configuration invariance | Composite invariance | | | Equality of variances | | | | Equal averages | | | | |
| Groups/Construct | | Original correlation | 5% | Supported partial measure invariance | Difference between original variances | 2.5% | 97.5% | Equal | Difference between original means | 2.5% | 97.5% | Equal | Supported total measure invariance |
| *Female managers—female employees* | | | | | | | | | | | | | |
| Existence | Yes | 0.812 | 0.300 | Yes | −0.032 | −0.523 | 0.356 | Yes | 0.236 | −0.354 | 0.360 | Yes | Yes |
| Accessibility | Yes | 0.989 | 0.823 | Yes | 0.112 | −0.593 | 0.472 | Yes | −0.120 | −0.375 | 0.362 | Yes | Yes |
| Emotional well-being | Yes | 0.995 | 0.993 | Yes | −0.133 | −0.478 | 0.344 | Yes | −0.270 | −0.361 | 0.333 | Yes | Yes |
| Physical well-being | Yes | 0.998 | 0.997 | Yes | −0.336 | −0.521 | 0.405 | Yes | −0.421 | −0.366 | 0.326 | No | No |
| Absenteeism | Yes | 0.988 | 0.309 | Yes | −1.850 | −3.293 | 2.105 | Yes | −0.029 | −0.246 | 0.451 | Yes | Yes |
| *Female managers—male managers* | | | | | | | | | | | | | |
| Existence | Yes | 0.853 | 0.329 | Yes | −0.164 | −0.454 | 0.368 | Yes | 0.277 | −0.396 | 0.458 | Yes | Yes |
| Accessibility | Yes | 0.998 | 0.979 | Yes | 0.048 | −0.676 | 0.659 | Yes | −0.307 | −0.428 | 0.471 | No | No |
| Emotional well-being | Yes | 0.989 | 0.991 | No | 0.182 | −0.683 | 0.611 | Yes | −0.578 | −0.412 | 0.414 | No | No |
| Physical well-being | Yes | 0.995 | 0.985 | Yes | 0.112 | −0.520 | 0.481 | Yes | −0.457 | −0.424 | 0.412 | Yes | Yes |
| Absenteeism | Yes | 0.800 | 0.280 | Yes | 0.967 | −1.387 | 1.278 | Yes | 0.257 | −0.434 | 0.433 | Yes | Yes |
| *Female managers—male employees* | | | | | | | | | | | | | |
| Existence | Yes | 0.923 | 0.611 | Yes | −0.200 | −0.442 | 0.327 | Yes | 0.376 | −0.383 | 0.366 | No | No |
| Accessibility | Yes | 0.998 | 0.952 | Yes | −0.144 | −0.552 | 0.396 | Yes | −0.406 | −0.335 | 0.369 | No | No |
| Emotional well-being | Yes | 0.998 | 0.995 | Yes | −0.197 | −0.481 | 0.359 | Yes | −0.380 | −0.343 | 0.359 | No | No |
| Physical well-being | Yes | 0.998 | 0.997 | Yes | −0.061 | −0.686 | 0.457 | Yes | −0.661 | −0.365 | 0.354 | No | No |
| Absenteeism | Yes | −0.336 | −0.272 | No | 0.235 | −1.384 | 0.967 | Yes | 0.205 | −0.336 | 0.382 | Yes | No |
| *Female employees—male managers* | | | | | | | | | | | | | |
| Existence | Yes | 0.967 | 0.868 | Yes | −0.198 | −0.326 | 0.389 | Yes | 0.010 | −0.288 | 0.296 | Yes | Yes |
| Accessibility | Yes | 0.988 | 0.930 | Yes | −0.063 | −0.438 | 0.492 | Yes | −0.217 | −0.321 | 0.296 | Yes | Yes |
| Emotional well-being | Yes | 0.998 | 0.996 | Yes | 0.320 | −0.296 | 0.398 | Yes | −0.269 | −0.322 | 0.302 | Yes | Yes |
| Physical well-being | Yes | 0.997 | 0.997 | Yes | 0.477 | −0.363 | 0.472 | No | 0.038 | −0.312 | 0.301 | Yes | No |
| Absenteeism | Yes | 0.962 | 0.453 | Yes | 2.324 | −1.734 | 2.960 | Yes | 0.103 | −0.311 | 0.220 | Yes | Yes |
| *Female employees—male employees* | | | | | | | | | | | | | |
| Existence | Yes | 0.984 | 0.966 | Yes | −0.094 | −0.215 | 0.217 | Yes | 0.117 | −0.194 | 0.179 | Yes | Yes |
| Accessibility | Yes | 0.986 | 0.987 | Yes | −0.250 | −0.278 | 0.249 | Yes | −0.326 | −0.194 | 0.184 | No | No |
| Emotional well-being | Yes | 1.000 | 0.999 | Yes | −0.064 | −0.222 | 0.215 | Yes | −0.117 | −0.189 | 0.167 | Yes | Yes |
| Physical well-being | Yes | 1.000 | 0.999 | Yes | 0.272 | −0.290 | 0.273 | Yes | −0.177 | −0.175 | 0.180 | No | No |
| Absenteeism | Yes | −0.178 | 0.406 | No | 2.340 | −2.349 | 2.415 | Yes | 0.143 | −0.181 | 0.194 | Yes | No |
| *Male managers—male employees* | | | | | | | | | | | | | |
| Existence | Yes | 0.986 | 0.894 | Yes | 0.069 | −0.355 | 0.264 | Yes | 0.130 | −0.300 | 0.306 | Yes | Yes |
| Accessibility | Yes | 0.993 | 0.971 | Yes | −0.189 | −0.421 | 0.357 | Yes | −0.134 | −0.286 | 0.304 | Yes | Yes |
| Emotional well-being | Yes | 0.998 | 0.997 | Yes | −0.390 | −0.439 | 0.319 | Yes | 0.143 | −0.311 | 0.292 | Yes | Yes |
| Physical well-being | Yes | 0.998 | 0.997 | Yes | −0.186 | −0.580 | 0.436 | Yes | −0.235 | −0.277 | 0.288 | Yes | Yes |
| Absenteeism | Yes | 0.362 | −0.533 | Yes | −0.526 | −1.139 | 1.087 | Yes | −0.132 | −0.270 | 0.334 | Yes | Yes |

Table 10. Multi-group analysis based on the permutations test.

| Groups/Direct Effects | Group 1 | | | Group 2 | | | Permutation | Significance |
|---|---|---|---|---|---|---|---|---|
| | $R^2$ | Direct Effect | *p*-Value | $R^2$ | Direct Effect | *p*-Value | *p*-Value | |
| *Female managers—female employees* | | | | | | | | |
| Absenteeism | 0.239 | | | 0.088 | | | | |
| Existence | | −0.250 | 0.202 | | −0.085 | 0.071 | 0.573 | No |
| Accessibility | | −0.459 | 0.005 | | −0.123 | 0.029 | 0.136 | No |
| Physical well-being | | 0.342 | 0.050 | | −0.202 | 0.017 | 0.090 | No |
| Emotional well-being | 0.244 | | | 0.183 | | | | |
| Existence | | −0.286 | 0.206 | | 0.290 | 0.000 | 0.019 | Yes |
| Accessibility | | 0.319 | 0.026 | | 0.262 | 0.000 | 0.758 | No |
| Physical well-being | 0.555 | | | 0.489 | | | | |
| Emotional well-being | | 0.745 | 0.000 | | 0.699 | 0.000 | 0.664 | No |
| *Male managers—male employees* | | | | | | | | |
| Absenteeism | 0.087 | | | 0.024 | | | | |
| Existence | | −0.090 | 0.352 | | 0.140 | 0.185 | 0.517 | No |
| Accessibility | | −0.113 | 0.312 | | −0.095 | 0.243 | 0.965 | No |
| Physical well-being | | −0.192 | 0.233 | | −0.056 | 0.292 | 0.581 | No |
| Emotional well-being | 0.399 | | | 0.207 | | | | |
| Existence | | 0.392 | 0.000 | | 0.203 | 0.001 | 0.184 | No |
| Accessibility | | 0.450 | 0.000 | | 0.361 | 0.000 | 0.546 | No |
| Physical well-being | 0.399 | | | 0.488 | | | | |
| Emotional well-being | | 0.632 | 0.000 | | 0.699 | 0.000 | 0.371 | No |

## 5. Discussion

The results we obtained have theoretical and practical implications. This work reinforces and updates the evidence in the literature on the negative relationship between WFP and absenteeism [25–27]. But, unlike in the previous studies in the literature, in our model we separately analysed the accessibility and the existence of the WFP in order to study this relationship. Our results indicate that the mere existence of WFP does not contribute to a decrease in absenteeism. On the contrary, the accessibility of WFP by workers can contribute to reducing absenteeism rates. Therefore, if managers want to reduce absenteeism, they should encourage WFP to be accessible to workers without reprisals of any kind. Their career opportunities and financial incentives should not be threatened. Furthermore, the organisation should express its support for the reconciliation of family life and personal life with work life. In this sense, the organisation should engage in the enhancement of cultural acceptance regarding family and work conciliation that would prevent social sanctions (that would imply the disapproval of co-workers and supervisors).

The results of this study also show the positive relationship between the existence and accessibility of WFP and the emotional well-being of workers. This relationship is consistent with previous studies in the literature that focus on the effects of WFP [22,23,52], but our findings provide a value added to this relationship, since we state that while the existence of WFP is important for emotional well-being, it is even more important that these WFP be accessible to workers.

We can confirm that, subsequently, there is a significant relationship between the emotional well-being and the physical well-being of workers [62], and that these two variables act as mediators between WFP and absenteeism. In this sense, our findings are consistent with previous findings in the literature that establish a negative relationship between well-being and absenteeism [103–105]. In practice, these relationships show that the lack of adequate WFP can cause work-family conflicts, thus reducing emotional well-being. This could generate job stress [106] which could, for example, subsequently lead to problems in the health of workers, thus ultimately increasing absenteeism. Therefore, managers should consider that workers with work-family conflicts not only experience less emotional well-being, but they may also experience health problems. According to the previous arguments, absenteeism can increase not only due to the difficulties that workers have in reconciling family and work, but also because they may subsequently suffer from stress, demotivation, and health problems.

In addition, demotivation and stress generated by inadequate WFP can reduce worker performance [22,23,51]. Whether workers are absent without the support and understanding of their organisation and superiors and without adequate conciliation measures, or are not absent but suffer from unresolved work-family conflicts, this could reduce their performance [22,23,51]. The cause would be that they could suffer high levels of stress and lower motivation as a result of their emotional distress or even health problems (physical distress). On the other hand, the well-being generated by WFP, in addition to influencing worker performance, can also increase worker commitment to the organisation and increase worker motivation and participation. Another advantage to be considered by managers would be that while some WFP, such as family leave, benefit the worker but not directly benefit the organisation, which has to continue to pay the worker's salary, there are other WFP that generate mutual benefits (see Table 1), such as unpaid leave, which do not represent an economic burden for the organisation, and flexibility in the workplace. Therefore, managers should be aware that the work-family balance of their workers provides the workers with additional benefits and reduces certain inconveniences for the organisation, beyond the reduction of absenteeism. Regarding the moderating variables of gender in the hierarchical position, we observed significant differences between the existence of WFP and emotional well-being between the groups of "female managers" and "female employees" (see Figure 1). These differences had not been observed in the previous studies in the literature, thus they generate added value in our work. However, this finding is consistent with the studies in the literature that note the special difficulty that female managers have in reconciling family life and work life [79–81], but these two groups do not show significant differences in the other relationships of the analysed model. On the other hand, the groups of "male managers" and "male employees" do not show significant differences in any relationship concerning the theoretical model.

## 6. Conclusions

The social and economic impact of absenteeism from work causes great concern to employers and academics [8]. The lack of physical well-being, emotional well-being, and reconciliation of family life and work life are among the main causes of absenteeism [11–14]. Reducing absenteeism requires organisations to implement appropriate WFP [14,19–21] and to consider the mediator role that could best promote well-being. Considering these arguments, this work has analysed the impact of the existence and accessibility of WFP on absenteeism, and the mediating role of emotional well-being and physical well-being in this relationship. Furthermore, the model includes the hierarchical position and gender as moderating variables. To test the proposed hypotheses, we applied a structural equation model based on the PLS-SEM approach to a sample of employees in the Spanish service sector.

Our findings show that the existence of WFP has no direct effect on absenteeism but, on the other hand, the accessibility of WFP does have a direct effect on absenteeism. Furthermore, both the existence and accessibility of WFP have positive direct effects on emotional well-being. In turn, emotional well-being has a positive relationship with physical well-being, and these two variables act as mediators between WFP and absenteeism. However, considering the moderation of the hierarchical position and gender in the model, we confirm significant differences between the groups of female managers and female employees in the impact that the existence of WFP has on emotional well-being.

Regarding the added value of this research, in our model we separately analysed the accessibility and existence of WFP, which is a different approach compared to the previous research in the literature. In addition, it is also novel to study the mediating role of well-being in the WFP-absenteeism relationship, and to do so while differentiating emotional well-being from physical well-being. On the other hand, we confirm the special difficulty of female managers in reconciling family life and work life.

Considering the practical implications of this work, according to our results, workers should perceive that WFP are accessible without subsequent retaliation of any kind. In this way, the WFP will contribute to the reduction of absenteeism, both directly and indirectly, due to the positive effect they have on the emotional and physical well-being of the workers.

Among the limitations of this study, it is noteworthy that WFP can only partially explain the emotional well-being of workers, and there are other aspects in organisations that influence this well-being. For example, Cheung, Lun, & Cheung [107] attribute the emotional well-being of workers to the quality of their jobs they undertake and their motivation for external recognition in the case of a job that is done well. Likewise, the WFP and the well-being that is generated explain only a part of the causes of absenteeism. An interesting future line of research would be to extend this study with other determinants of absenteeism. Another limitation is that the sample was obtained from the Spanish service sector. Future replications of this study in other sectors and countries, with different labour regulations, cultures, and degrees of economic development, would improve the generalisability of the results. The reduced number of cases of female managers that were used in the comparison between groups is also a limitation of this study. This limitation does not undermine the value of the findings. It is a statistically adequate number, but the comparison between groups would require verifying the results with groups of similar size [108]. Another interesting line of research would be to study the determinants of turnover using the theoretical model of this work, since the research in the literature identifies some common determinants for turnover and absenteeism [109,110]. Finally, the well-being that is generated by WFP could also increase worker commitment, motivation, participation, and performance [22,23,51]. In this sense, it would be interesting to analyse the mediating role of well-being between WFP and these variables.

**Author Contributions:** The authors contributed equally to curating and analysing the data and writing the paper. All authors have read and agreed to the published version of the manuscript.

**Funding:** This research was partially funded by INDESS of the Universidad de Cádiz, and the APC was partially funded by Department of Business Organization of the Universidad de Cádiz.

**Conflicts of Interest:** The authors declare there to be no conflict of interest.